\documentstyle[aps,epsf]{revtex}

\def\myfigure#1#2{{\leftskip=0.10753\textwidth \rightskip\leftskip\small
\begin{figure}\baselineskip=14pt plus 2pt minus 1pt
\centerline{#1}\nobreak\smallskip\nobreak #2\end{figure}}}
\global\firstfigfalse

\begin{document}
\newtheorem{guess}{Proposition }[section]
\newtheorem{theorem}[guess]{Theorem}
\newtheorem{lemma}[guess]{Lemma}
\newtheorem{corollary}[guess]{Corollary}
\def \ja {\vrule height 3mm width 3mm}

\title{ The  Classification of Surfaces}
\author{Elizabeth Gasparim}  
\address{Departamento de Matem\'atica, Universidade Federal de 
Pernambuco \\Cidade Universit\'aria, Recife, PE, BRASIL, 50760-901 \\
e-mail:gasparim@dmat.ufpe.br \thanks{Partially supported by a research 
grant from CNPQ (Brasil)}}
\author{Pushan Majumdar}
\address{Institute of Mathematical Sciences,C.I.T campus Taramani. Madras 
600-113. India. \\ e-mail:pushan@imsc.ernet.in} 
\maketitle

\begin{abstract}
This is an expository paper which  presents the holomorphic classification 
of rational complex 
surfaces from a simple and intuitive point of view, which is 
not found in the literature. 
Our approach is to compare this classification 
with the topological classification 
of real surfaces.
\end{abstract}

\section{Introduction}

 Our aim here is to present the holomorphic 
classification of complex rational 
surfaces in a simple way. We base ourselves on the fact that the 
topological classification of real surfaces is very intuitive as one can 
see by following our pictures. One should keep in mind that a real 
surface is a two dimensional object while a rational surface is a four 
dimensional object. Hence, we can follow the classification of real 
surfaces by drawing pictures and use the intuition built in this case to 
understand the classification of rational surfaces, for which we are no 
longer able to draw pictures. When classifying real surfaces it is 
natural to take the following steps.
\begin{enumerate}
\item Basic surfaces : where we give the first elementary examples of 
real surfaces.
\item Two types of surfaces : where we show that real surfaces are 
naturally divided into two categories, namely orientable and non-orientable.
\item Constructing new surfaces out of basic ones : where we define the 
operation of connected sum and show how to build new examples of real 
surfaces.
\item The classification theorem : that says that all real surfaces can 
be obtained out of the basic ones by means of the connected sum operation.
\end{enumerate}

We give an outline of the proof of the classification theorem for real 
surfaces.

Then we present the classification of rational surfaces following 
exactly the same four steps as above with the appropriate modifications.
\begin{enumerate} 
\item Basic surfaces : where we give the first elementary examples of 
complex surfaces.  
\item Two types of surfaces : where we show that complex surfaces are 
naturally divided into two categories, namely rational and 
irrational.  
\item Constructing new surfaces out of basic ones : where we define the 
operation of blowing up and show how to build new examples of complex
surfaces.  
\item The classification theorem : that says that all rational surfaces 
can be obtained out of the basic ones by means of the blowing up
operation. 
\end{enumerate}

 We give an informal presentation and  indicate references 
for detailed proofs. We only assume some elementary 
knowledge of smooth manifolds and some linear algebra.
By a real surface we mean a smooth manifold of two real dimensions, 
which is connected and compact.
 By a complex surface we mean a complex smooth manifold 
of two complex dimensions (hence four real dimensions) which is connected 
and compact. 

\section{The Topology of Real Surfaces}

\noindent {\bf 1. Basic surfaces.}
We begin with examples of real surfaces.
\begin{itemize}
\item the sphere $S^2$

\item  the torus $T$ ( which looks like a doughnut)

\item  the n-torus $nT$ 
( which looks like a doughnut with $n$ holes )

\begin{center}{\bf The orientable surfaces}\end{center}
\myfigure{\epsfysize 2.4in\epsfbox{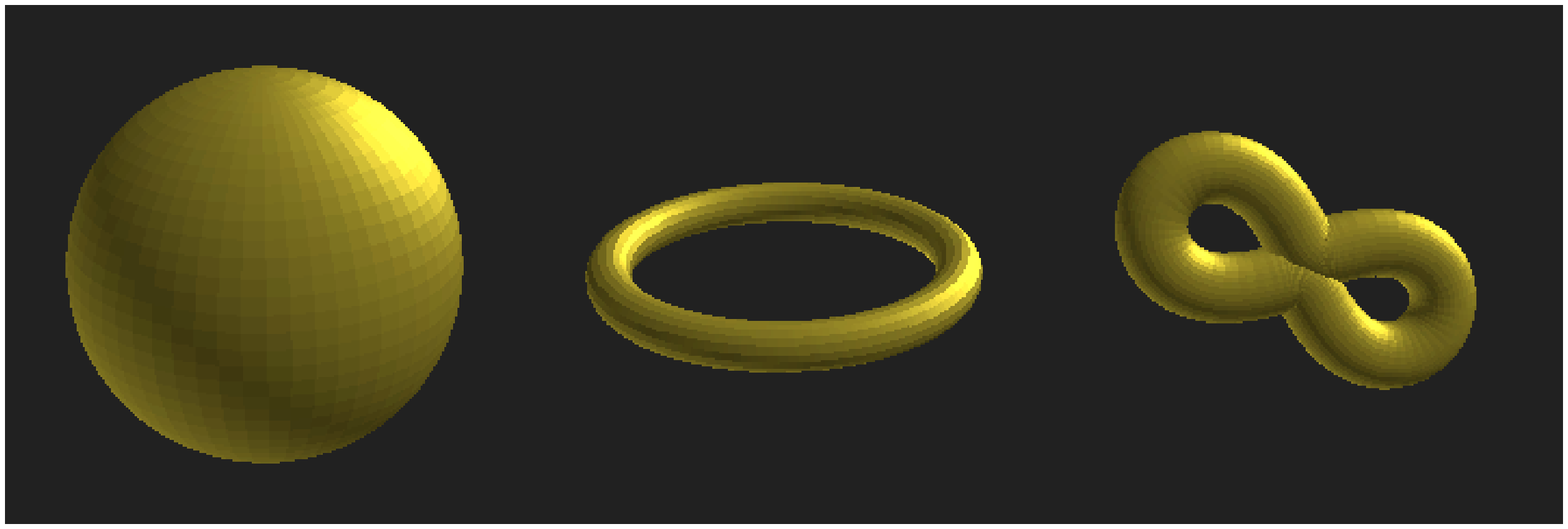}}{\hspace*{-1.5 cm}
fig. 1. The Sphere \hfill fig. 2. The Torus \hfill fig. 3. The 2-torus}

\item the real projective plane 
${\bf RP}^2 \sim {{\bf R}^3 - \{ 0 \}
 \over z \sim \lambda z },$ for $\lambda \in {\bf R} - \{0 \}.$
\item the Klein bottle
 $K,$ which can be viewed as a twisted torus, constructed
by gluing the two ends of a cylinder with a twist.

\begin{center}{\bf A Non-orientable surface}\end{center}
\myfigure{\epsfysize 2.4in \epsfbox{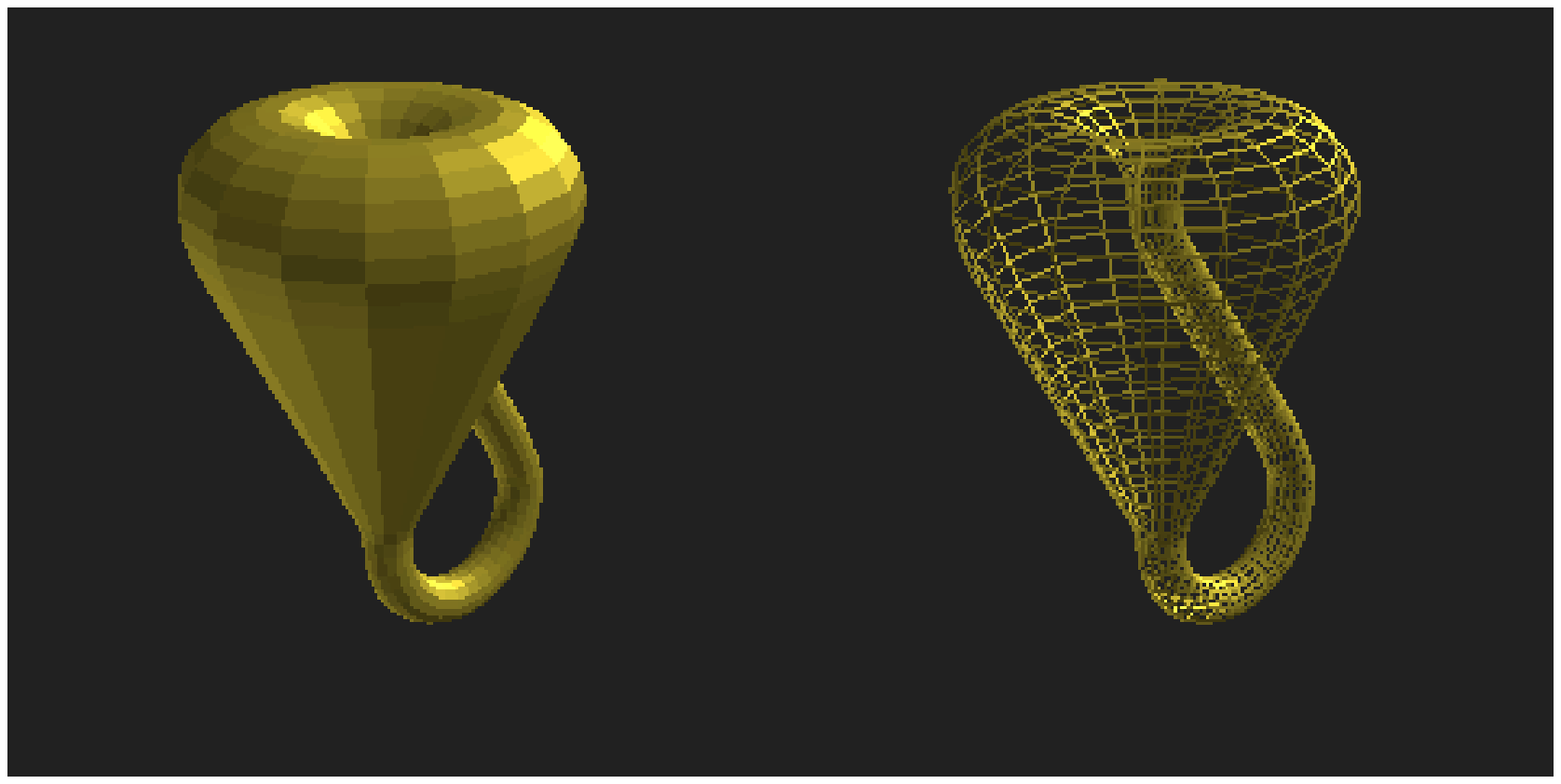}}
{\begin{center} fig. 4. The Klein Bottle \end{center}}

\end{itemize}

\noindent {\bf 2. Two types of surfaces.}

\vspace {3 mm}

There is a natural division in the 
classification of real surfaces,  given by orientability.
Orientable surfaces are intuitively the ones that have two sides, that is 
inside and outside, like $S^2,$  $T,$ and $nT.$ Orientable 
real surfaces are the so called Riemann surfaces.
The non-orientable surfaces are the ones containing a 
M\"obius band, which therefore don't have outside and inside, like 
${\bf RP}^2$ and $K.$\\

\noindent{\bf 3. Constructing new surfaces out of the basic ones.}

\vspace {3 mm}

When one wants to study the topology
of real surfaces, there is a basic operation, which allows us to construct 
new surfaces out of  old ones. This operation is called 
connected sum (represented by \#),
 and works as follows. Given two surfaces 
$S_1$  and $S_2,$ cut out open  discs
 $D_1$ in $S_1$ and  $D_2$ in $S_2$
and then glue the two surfaces $S_1 - D_1$ 
and $S_2 - D_2 $  by identifying the boundaries of 
$D_1$ and $D_2.$ 
Here are some examples.

\begin{itemize}
\item $X$ \#  $S^2$ = $X$ for any real surface $X,$ that is, the sphere 
works as an identity for the \# operation
\item $T$ \# $T$  = $2T,$ that is, the 2-torus is obtained by gluing 
together two tori (Intuitively, just put together two 
doughnuts one after the other and you get a 2-torus.)

\myfigure{\epsfysize 2.4in\epsfbox{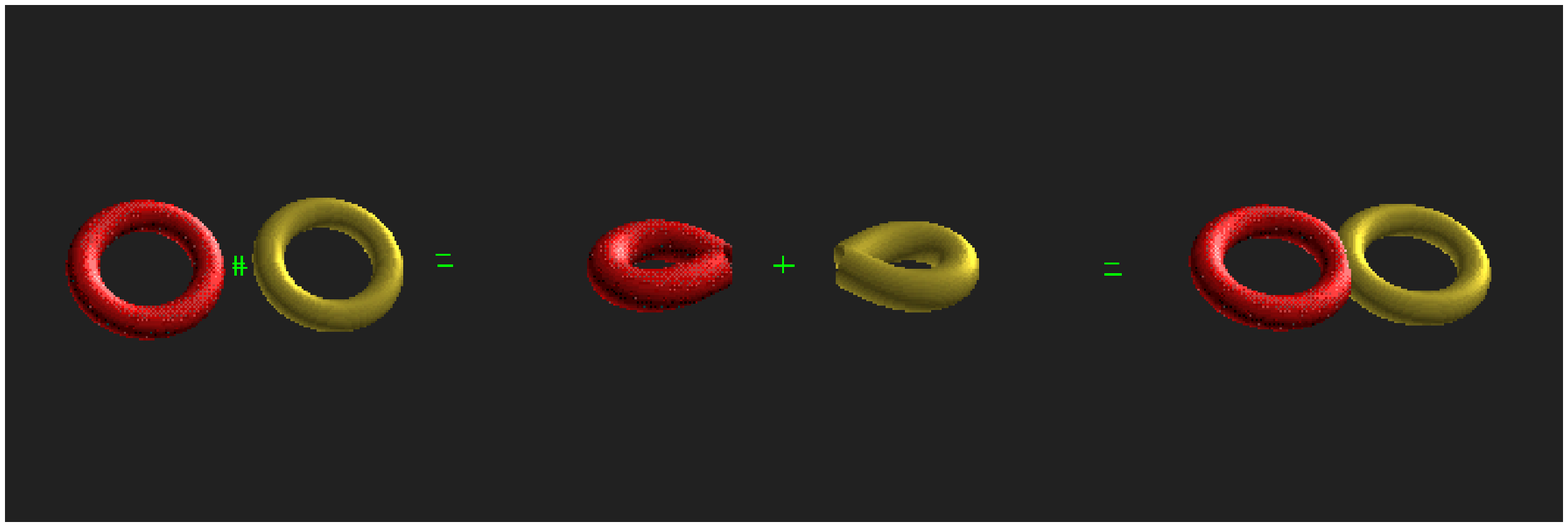}}{
\begin{center}fig. 5. The Connected Sum of Two Tori\end{center}} 

\item $nT$ \# $mT $ = $(n+m)T,$ which is a generalization of the 
previous example.
\item ${\bf RP}^2$ \#  ${\bf RP}^2$ = $K.$ This case is not intuitively 
obvious, but is very easy to show by cuting and pasting. We give a proof 
of this in section III.

\end{itemize}

\noindent{\bf 4. The classification theorem.}

\vspace{ 3 mm}

Once we have the basic surfaces and the
operation for constructing new surfaces 
out of old ones, we can build all 
real surfaces.
The classical structure theorem is
the following (see [2]).

\begin{theorem} Every compact connected real surface 
is obtained from either the sphere $S^2$ or the torus $T$
or  the real projective plane ${\bf RP}^2$ by connected sums.
\end{theorem}

\section{The proof of the structure theorem}

In this section we outline the proof of the classification theorem for 
real surfaces. For a detailed proof see [2]. By the very definition of 
the connected sum operation, we know that performing connected 
sums on real surfaces, we always obtain a real surface. However, it is 
not obvious that all real surfaces are obtained by connected sums 
starting only with the sphere, the torus and the real projective plane. 
So in this section, we will give an idea of how to show this.

To prove this theorem one very useful tool is to construct surfaces by 
identifying sides of polygons. Here one should keep in mind that we want 
a topological classification, so that an object is considered the same 
as a continuous deformation of it which can be reversed. For example, 
from a topological point of view a disc and a square are considered 
the same topological object, since each can be continuously deformed to 
the other.

Here are some examples of representing real surfaces by polygons with 
sides identified. We will name our polygons according to their sides 
read counterclockwise.

\begin{center}{\bf The Polygon Representations}\end{center}

\myfigure{\epsfxsize 6.5in\epsfbox{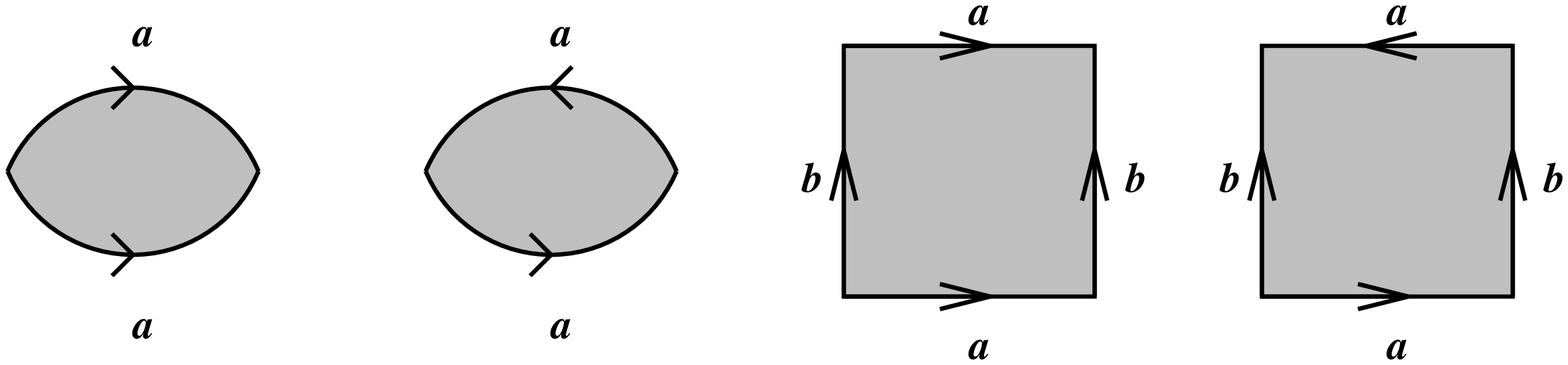}}
{\begin{flushleft}\makebox[7in]{ 
fig.6 The Sphere \hspace*{.9cm}fig.7 The Real Projective Plane  
\hspace*{1.2cm} fig.8 The Torus \hspace*{1.6cm}fig.9 The Klein Bottle}
\end{flushleft}}

Now we also need to represent connected sums using this method of 
polygons. This is easily seen by an example.

\myfigure{\epsfxsize 5in\epsfbox{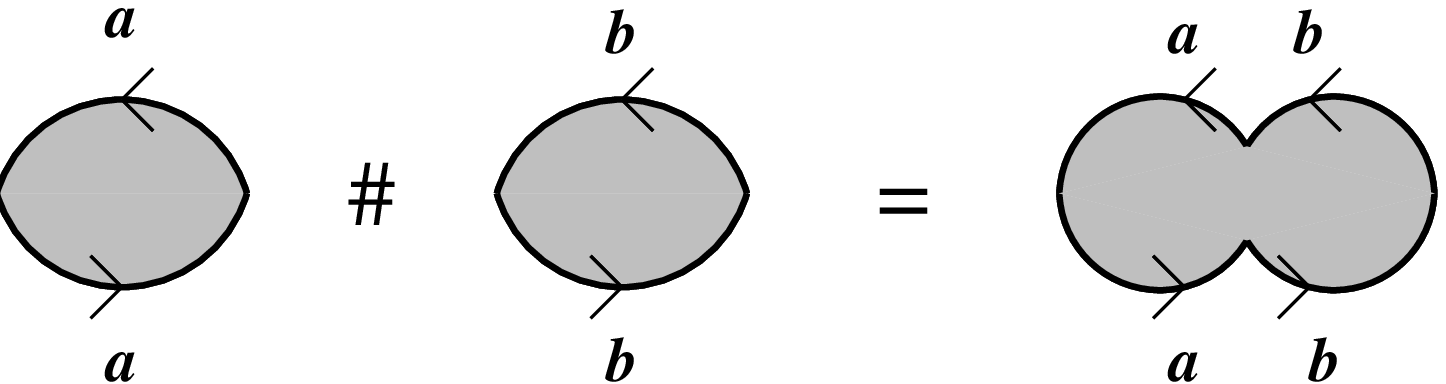}}{\begin{center}
fig.10 Schematic representation of the sum of two real projective planes 
\end{center}}

We can represent any of our real surfaces by a polygon with sides 
identified. However this representation is not unique. To understand this 
look at the following sequences which shows that $RP^{2}\# RP^{2}$ equals 
the Klein bottle. Let us simply continue from the previous sequence of 
pictures. We cut our polygon with sides $aabb$ through a diagonal side $c$.

\myfigure{\epsfxsize 4in\epsfbox{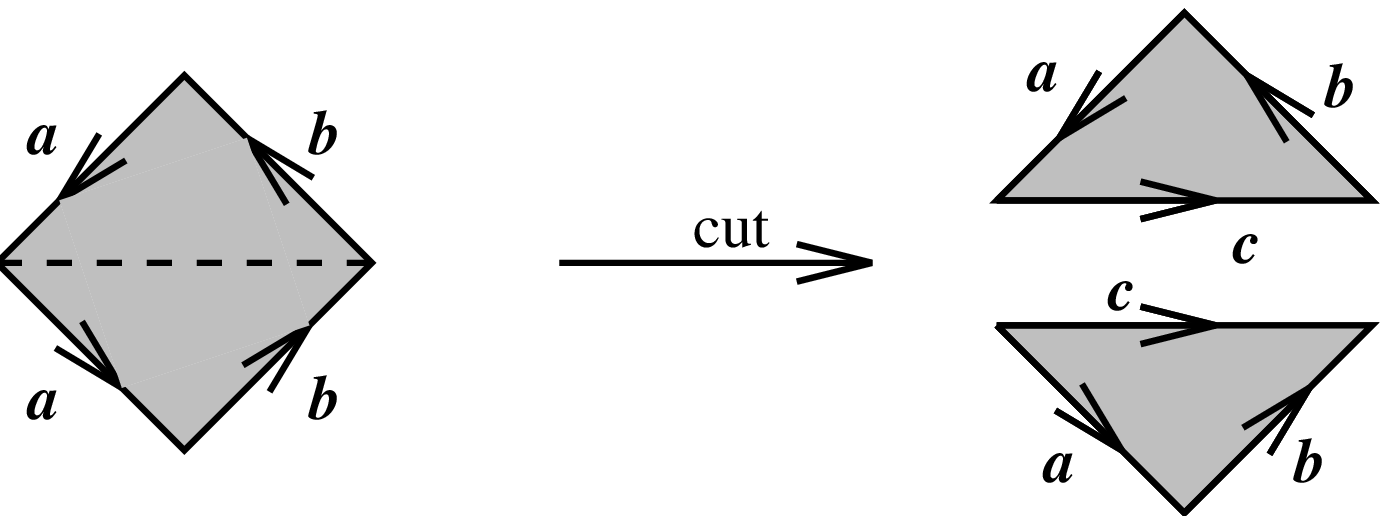}}{\begin{center} fig.11 The Cut
\end{center}}

Then we glue the two triangles back together by their $b$ sides to obtain 
the sequence of sides $aca^{-1}c$ which is the Klein bottle.

\myfigure{\epsfxsize 5in\epsfbox{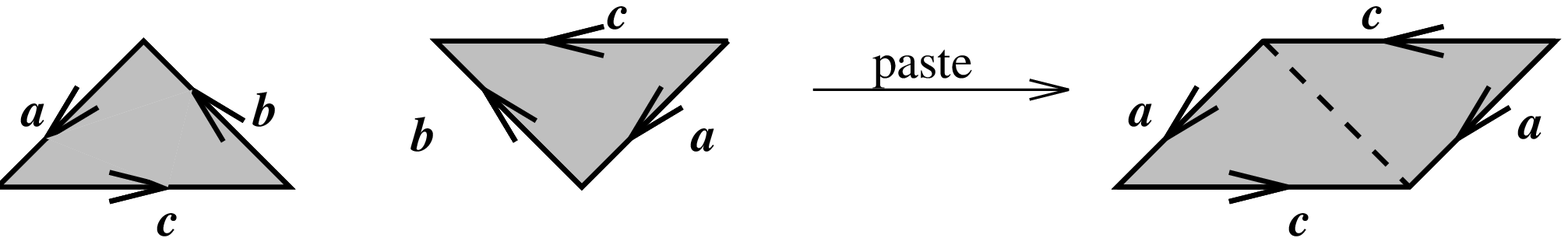}}{\begin{center} fig.12 The 
Paste \end{center}}

Now we are ready to understand the proof of the structure theorem of 
real surfaces. The proof goes as follows. Every real surface is obtained 
by identifying sides of a polygon. This is intuitively believable. The 
rigorous mathematical reason is that every surface has a triangulation.

Then to prove the theorem one wants to see that one can choose a 
representation which shows the correspondence with a connected sum of 
tori, spheres or real projective planes. First of all it is clear 
that each letter appears in two sides of the polygon since we want to 
obtain a closed surface. Hence for every surface we get a sequence of 
letters where each letter appears twice. We put the exponent $-1$ when 
a letter appears in a side pointing in the clockwise direction. Then we 
use the operations of cutting and pasting as we did in the above example 
to get the polygon to a nice form. One shows that cutting and pastings 
can be done to get the sequence of letters to be always of one of the 
types $aa$, $aa^{-1}$ or $aba^{-1}b^{-1}$ following each other any 
number of times. For example we can have $a_{1}b_{1}a_{1}^{-1}b_{1}^{-1}
a_{2}b_{2}a_{2}^{-1}b_{2}^{-1}\cdots a_{n}b_{n}a_{n}^{-1}b_{n}^{-1}$ 
which is a connected sum of n tori. Mathematically one gets a group 
generated by expressions of the form $aa$, $aa^{-1}$ and 
$aba^{-1}b^{-1}$. Each geometric property of a surface gets translated 
into a property of the group. For example, the fact that the sphere is 
an identity element for connected sums gets translated into the fact 
that $xaa^{-1}=x$ in the group language. The best way to understand 
this result is to draw polygons on pieces of papers and perform cutting 
and pasting until one gets to the form of connected sums of the basic 
elements.

Summing up, given a surface, one represents it by a polygon with sides 
to be identified. Then by cutting and pasting one can always get to a 
simple representation of the surface as a polygon containing only 
sequences of sides of the form $aa, aa^{-1}$ and $aba^{-1}b^{-1}$. These 
are exactly the expressions that correspond to $RP^{2}, S^{2}$ and $T$.

\section{The structure of rational surfaces}

Here we  imitate step by step the
classification of  real surfaces  presented in section II.
We begin with some examples.\\

\noindent {\bf 1. Basic surfaces.}
Here we give examples of complex surfaces.
\begin{itemize}
\item the 
complex projective plane ${\bf CP}^2$
\item the product of two 
projective lines   ${\bf CP}^1 \times {\bf CP}^1$
\item  The Hirzebruch surfaces $S_n.$
\end{itemize}

For the first two examples, recall that
${\bf CP}^n := {{\bf C}^{n+1} - \{ 0 \}
 \over z \sim \lambda z },$ for $\lambda \in {\bf C} - \{0 \}.$
To define the Hirzebruch surfaces,
we need the concept of a vector bundle.
Intuitively, a rank n vector bundle over 
a manifold $M$ is given by attaching to each point of 
the manifold a rank n vector space $F,$ which is called 
the fiber. In other words a
vector bundle over $M$ is 
a family of vector spaces parametrized 
by the points of $M.$  One very basic property
of vector bundles that one should keep 
in mind is called local triviality. This means that when 
we look at a small
 open set $U$ in the manifold $M, $ then the vector bundle
over $U$ 
is isomorphic to a trivial product $U \times F.$
Here we only need  vector bundles over 
${\bf CP}^1,$ and we will only define these. For more 
on vector bundles see [4].

First of all we choose charts for 
the complex projective line 
as follows: 
${\bf CP}^1 = U \cup V,$ where
$U \simeq V \simeq {\bf C}$ with intersection $U \cap V \simeq
 {\bf C} - \{0\},$
 and transition function $z \rightarrow z^{-1}.$
Topologically ${\bf CP}^1$ is just the sphere $S^2$ and 
one can look at the chart $U$ as covering the sphere minus the north pole,
while $V$ covers the sphere minus the south pole.

Now we  define the rank one complex vector bundle ${\cal O}(n)$
over ${\bf CP}^1$
(rank one complex means 
that the fibers will be copies of ${\bf C}$).
To define this bundle 
we start with two trivial products $U\times {\bf C}$ and 
$V \times {\bf C}.$
Then  we need  to give
a transition function  that  tells us how the fibers
get put together in the overlap of the charts.
For each n we give the function $z^{-n},$ which means that 
we identify these fibers by the 
vector space isomorphism 
taking $u$ to $z^{-n}u.$
Formally we define ${\cal O}(n)$
 by charts $U'  = U \times {\bf C}  \simeq {\bf C}^2,$
 $V'  = V \times {\bf C}  \simeq {\bf C}^2,$ with intersection 
$U' \cap V' \simeq {\bf  C} - \{0\} \times {\bf C},$ and 
transition function $ (z,u) \rightarrow (z^{-1}, z^{-n}u).$

Now we construct rank two vector bundles over 
${\bf CP}^1$ using the operations 
we know for vector spaces coming from basic linear algebra.
Recall that we have an operation of direct sum for 
vector spaces, which for vector spaces $V_1$ of rank $r_1$ 
 and $V_2$ of rank $r_2$
associates the vector space $V_1 \oplus V_2$ 
of rank $r_1 + r_2.$
Now starting with  two vector bundles $E_1$ and $E_2$ 
 over a manifold $M,$ we  construct 
a new bundle 
over $M$ called  the Whitney sum  $E_1 \oplus E_2,$
 whose fibers  are just the 
vector space sum of the fibers of $E_1$ and $E_2.$ That is
if $E_1$ has a fiber $V_1$ at the point $p$
and  $E_2$ has a fiber $V_2$ at the point $p$
then $E_1 \oplus E_2$ has fiber $V_1 \oplus V_2$ at the point $p.$
For example, taking  the Whitney sum of 
the  bundles  ${\cal O}(m)$  and $ {\cal O} (n)$
will give us a  rank two bundle over ${\bf CP}^1$ 
denoted by ${\cal O}(m) \oplus {\cal O} (n).$

Now, just as we constructed  ${\bf RP}^2$ by projectivizing 
${\bf R}^3$ and ${\bf CP}^2$ by projectivizing 
${\bf C}^3,$ it is possible  to projectivize any 
vector bundle $E,$ thus obtaining a bundle $P(E)$ whose fibers are 
projective spaces. Also here
 the idea is to perform the projectivization at each 
of the fibers.
If we start with the rank two bundles  ${\cal O}(n) \oplus {\cal O} (0)$
 then projectivizing we transform the fibers
into projective lines ${\bf CP}^1.$ Recall that
these are all bundles over  ${\bf CP}^1.$ Hence what we obtain 
is actually a family of 
 ${\bf CP}^1$'s parametrized by 
points of another copy of  ${\bf CP}^1.$ 
The reader should intuitively think of the 
number $n$ as a particular way 
of twisting such a family $n$ times when we go around the base.

\noindent We are now ready to  define the n-th Hirzebruch surface:
$$S_n = P( {\cal O}(n) \oplus {\cal O}(0)).$$

Each Hirzebruch surface
$S_n$ is a  complex surface. Hence now we 
have an infinite set of  complex surfaces, one for each $n.$
We remark also that the first Hirzebruch surface $S_0$ is simply 
the product ${\bf CP}^1 \times {\bf CP}^1,$
intuitively in this case we have a 
family of  ${\bf CP}^1$'s parameterized by a  ${\bf CP}^1$
with no twisting hence a  product.\\

\noindent{ \bf 2. Two types of surfaces.}

\vspace {4 mm}

Just as real surfaces were naturally divided into two categories,
also complex surfaces will be divided into two categories. 
However orientability does not make a division, because 
all complex surfaces are orientable.
There is  a natural division in the 
classification of complex surfaces is given by rationality.
Let us observe that all complex surfaces we mention here can be thought of
as lying inside some complex projective space. In a 
projective space it always make sense to 
write down quotients of homogeneous polynomials of 
the same degree. 
One then defines a rational map to be a map that 
is locally given (in each coordinate) by  quotients
of polynomials. We remark a very special property of rational 
maps that is the fact that a rational map need not be
defined over every point of its domain.
 Clearly when we define a map by quotients of polynomials,
then the map is not defined where the denominator vanishes.
In fact a rational map needs only to be defined over a 
dense open subset of its domain and two rational 
maps are considered equal if they coincide on a dense open set.

Rational surfaces are intuitively the ones that
are somehow similar to ${\bf CP}^2.$ By definition, a complex surface $S$
is rational if there is a rational map $\Phi: S \rightarrow {\bf CP}^2,$
whose inverse is also rational. Where two rational maps are 
called inverse to each other if their composition is the identity 
defined over a dense open set.
Obviously ${\bf CP}^2$ itself is rational.
 Also the Hirzebruch surfaces are 
 rational surfaces and these are in fact the 
simplest examples of rational surfaces.\\

\noindent{\bf 3. Constructing new surfaces out of basic ones.}

\vspace {4 mm}

When one wants to study all rational 
 surfaces, there is a basic operation, which allows us to construct 
new surfaces from old ones. This operation is called 
blow-up. 
When we start  with a complex surface $S$
and blow up a point in $S$, the result is another complex 
surface
$\widetilde{S}$ which is different from $S.$ 
In fact topologically, the new surface $\widetilde{S}$ 
is obtained from $S$ by performing a connected sum
of $S$ together with a copy of ${\bf CP}^2$
with the reversed orientation. The orientation problem is a technical detail
that we will explain after we define the blow-up.
But right now it is important to see 
that when we do a blow-up, then from a topological point of 
view we are doing exactly the same kind 
of operation that we did for real surfaces in 
section II.
Hence  some of the geometric intuition 
we got from the pictures  can be carried over to this 
section.
However, we  remark also that the 
blow-up operation 
is an analytic procedure in the 
sense that it takes us from 
a complex analytic manifold into another complex analytic 
manifold.
That means that the blow-up operation is much more powerful then simply
performing connected sums in a topological way as we did for real 
surfaces. We could phrase this as  {\em the blow-up 
is a complex  analytic way of performing connected sums}.

The blow-up operation is defined locally, in the sense  that 
it is done in a coordinate chart. That is, to blow-up a point 
$p$ in a surface $S,$ we take a coordinate chart around $p,$
taking $p$ to $0 \in {\bf C}^2.$ Then we blow up the origin in 
${\bf C}^2$ and construct the surface $\widetilde{S},$ which is 
isomorphic to  $S$ everywhere except at the point $p,$ which gets replaced by
a ${\bf CP}^1.$ Intuitively one picks a point and blows it up (in the usual
sense of the word, i.e. to explode) to a 
projective line. The result is that the lines passing through p become 
disjoint. The next following  picture represents
the blow-up of ${\bf R}^2$ at the origin.
Note that after the blowup we 
get a mobius band as the top and bottom edges of the strip are identified 
after giving a twist. A good way to understand this picture is to think 
of the blow-up as placing a line $L$ through 
origin at the height $\theta$ where $\theta$ is the angle 
of $L.$ Then the line corresponding to $2\pi$ 
is to be identified to the line corresponding to zero, since they 
coincide in ${\bf R}^2.$ 

\myfigure{\epsfxsize 6in\epsfbox{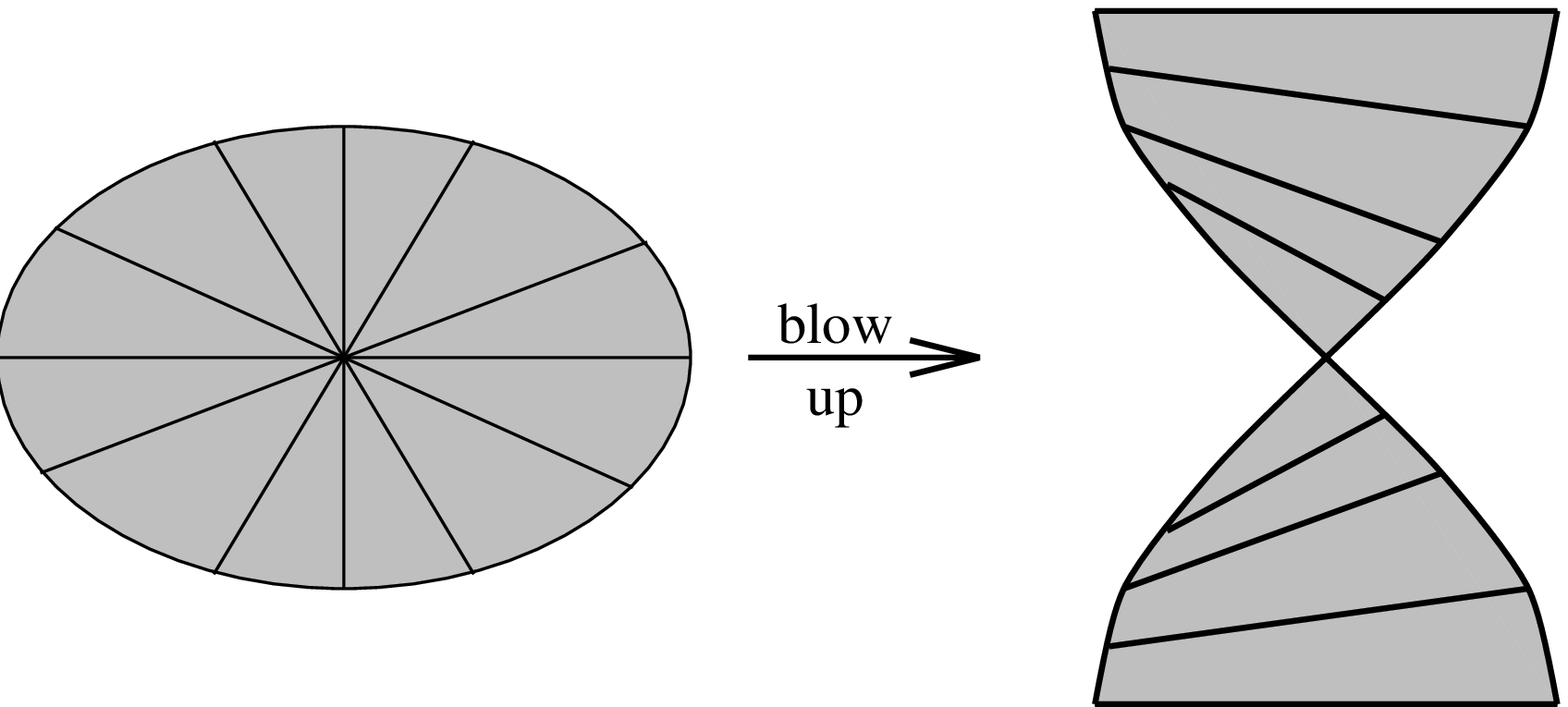}}{\begin{center} fig.13 The 
blow up of a point \cite{H} \end{center}}

Let us now write the formal  definition of  
the blow up of ${\bf C}^2$ at the origin, denoted by $\widetilde{{\bf 
C}^2},$ which is:
$$\widetilde{{\bf C}^2} = \{(z,l) \in {\bf C}^2 \times {\bf CP}^1 / \,  z \in l\}.$$

Notice that if $z \in {\bf C} $ is nonzero, then 
$z$ determines a unique line $l$ 
   namely the line through the origin in ${\bf C}^2$
passing by $z.$  Hence, over  $z \neq  0$ 
$\widetilde{{\bf C}^2}$ is isomorphic to ${\bf C}^2,$
which is what we should expect since points outside of the origin 
are not blown-up.
However, for $z = 0$ we have that $(0,l) \in \widetilde{{\bf C}^2}$
for every $l.$ This gives a copy of ${\bf CP}^1$  
which lies over the origin, called the exceptional divisor.
This exceptional divisor ``separates'' the lines through the origin 
turning them into disjoint lines. 
One very nice exercise for the reader would be to show that 
$\widetilde {\bf C}^2 $ is the line bundle
$ {\cal O}(-1)$ over ${\bf CP}^1.$ But for a first approach
 the reader may just take this as a result.

A copy of   ${\bf CP}^1$ inside a surface is called a line.
The exceptional divisor is a special kind of 
 line.
For instance it  is different from any line
contained in   ${\bf CP}^2.$ 
We  now assign numbers to lines according to 
the form of their neighborhoods, this number is 
know as the self-intersection number
of the line.
A line  ${\bf CP}^1$ inside  ${\bf CP}^2$ will be 
our standard line. To it we assign the number 1.
To the exceptional divisor we will assign the number
-1 (the technical reason for this is 
the exercise we just left to the reader).

For a line which has a neighborhood 
isomorphic to 
${\cal O}(n)$ we assign the number $n.$
These numbers are the analogous to the 
directions for the arrows we had when dealing with 
real surfaces. Only in that case only two direction were 
possible, so an arrow was enough. Now we have ``twists''
and we label the lines according to the twists in their neighborhoods.
A negative sign can be thought of  as meaning twist in  negative direction.
It is this negative sign that accounts for the 
 reversed orientation when we talked about the topology 
of the blow-up viewed as a connected sum.

An interesting example is to  start with 
${\bf CP}^2$ and blow up two points $p$ and $q,$ and then we blow down the 
line  passing  by $p$ and $q.$  Then, the resulting 
surface is ${\bf CP}^1 \times {\bf CP}^1.$  This is a lot more 
technical, we refer the reader to [1] for a proof.

Not only the blow-up is  topologically what we 
would expect for a generalization of the operation we had for real surfaces 
and even more, an analogous structure theorem holds.\\

\noindent{\bf 4. The classification theorem.}\\

Once we have the basic surfaces and the
operation for constructing new surfaces 
out of old ones, we can build all 
rational surfaces.
The classical structure theorem is
the following (see [1]).

\begin{theorem} Every compact connected rational surface 
is obtained from either the complex 
 projective plane ${\bf CP}^2$    or from a 
Hirzebruch surface $S_n$ by blowing up points.
\end{theorem}

A complete proof of this theorem would require that we develop 
quite a bit of theory.
Hence we give only some informal ideas.
We start with  a rational surface
and look for  special lines inside the surface.
Then there are two possibilities either we have lines with associated number 
-1 or we do not have any such lines. 
Suppose $S$ is a surface containing a -1 type line $l.$
Then we can eliminate this line $l$ by
performing what is called a blow-down, that is exactly the 
inverse operation of a blow-up. The blow-down of $l$ 
contracts the line into a point $p$ giving a new surface $S'$
(so that the blow-up of $S'$ at $p$ equals $S)$.
Then if $S'$ has no more -1 type lines we are done. Otherwise 
we perform another blow-down on $S'.$ 
It is true that after finitely many blowing downs we 
arrive at a surface containing no more -1 type lines.
Then there are two possibilities, either the 
new surface has only type 1 lines  and  in this case 
it is ${\bf CP}^2$ or it has some line of type $-n,$
in which case it is 
a Hirzebruch surface $S_n.$


\begin{thebibliography}{99}

\bibitem{GH} Griffiths, P.  and  Harris, J. {\em Principles of Algebraic  Geometry. John Wiley and  Sons, Inc.(1978) }
\bibitem{H} Harris, J. {\em Algebraic Geometry  A First Course, Graduate Texts in Mathematics 133 , Springer Verlag (1992)}
\bibitem{MA} Massey, W. S. {\em  Algebraic Topology, An Introduction. Graduate Texts in Mathematics 56 , Springer Verlag (1977)}
\bibitem{ST} Steenrod, N. {\em  The Topology of Fiber Bundles, Princeton University Press (1951)}


\end{thebibliography}
\end{document}